Submitted to **ADVANCED MATERIALS**

# Flexible Metal Oxide/Graphene Oxide Hybrid Neuromorphic Devices on Flexible Conducting Graphene Substrates

By Chang Jin Wan [1, 2], Wei Wang [2], Li Qiang Zhu [2],Yang Hui Liu [2], Ping Feng[1], Zhao Ping Liu [2], Yi Shi [1*], and Qing Wan[1, 2, *]

C. J. Wan, Prof. P. Feng, Prof. Y. Shi, Prof. Q. Wan
School of Electronic Science & Engineering, and Collaborative Innovation Center of Advanced Microstructures, NanjingUniversity, Nanjing 210093, China
E-mail: wanqing@nju.edu.cn; yshi@nju.edu.cn

Prof. W. Wang, Prof. L. Q. Zhu, L. Y Hui, Prof. Z. P. Liu
Ningbo Institute of Material Technology and Engineering, ChineseAcademy of Sciences, Ningbo 315201, China.

Human brain has a massively parallel and reconfigurable network with~$10^{11}$ neurons and ~$10^{15}$ synapses, and pyramidal neuron that receives spike inputs from thousands of synapses distributed across dendritic trees is often considered to be the computational engine of the brain[1]. Synaptic plasticity is the biological process by which specific patterns of synaptic activity result in changes in synaptic strength and is thought to contribute to learning and memory [2]. The idea of building brain-inspired adaptive artificial devices has been around for decades. At the outset, two-terminal devices such as memristors, phase change memory, and atom switches, etc, have been explored as the building blocks of neuromorphic systems [3-12]. More recently, three-terminal neuromorphic devices have been demonstrated in the pursuit of the synaptic plasticity and computation in a single device with scalable and low-energy consumption [13-23]. The reported devices reveal alternative potentials in acting more than a weight tunable connection, which can perform signal processing/computing serving as synaptic filter, integrator, etc, in neuromorphic circuits [17, 18]. For example, synaptic filtering functions based on short-term plasticity was successfully mimicked in individual oxide-based synaptic transistor, and the frequency-dependent gain is as high as 6.0 at 50 Hz [17]. In vivo, synapses utilize a wealth of nonlinear mechanisms to transform synaptic input, which underlie a range of arithmetic operations on signals [24]. These functions achieved in individual three-terminal neuromorphic device could be further applied to reproduce intelligent behaviors of biological nerve systems such as collision avoidance, pattern recognition, and sensory processing.

Mechanical flexibility has been regarded as great traction in many new-concept electronic systems[25-27]. For example, ultrathin and flexible silicon nanomembrane transistorarray were proposed to record spatial properties of cat brain activity in vivo, which makes a high resolution brain-machine interface devices possible for the diagnostic and therapeutic applications[28]. Flexible neuromorphic devices with mechanical flexibility and neuromorphic computation ability thus are of great significance for a wide range of applications such as brain-computer interface, conformable bionic sensors, and neuromorphic systems [23, 28]. Graphene is a very promising material for flexible devices because of its high mechanical property, high chemical stability and high stretchability *etc.* For example, graphene film deposited by chemical vapor deposition (CVD) method presents a low sheet resistance of few hundreds $\Omega \cdot \square^{-1}$ or less and a high transmission of over 97.4%, which has been also demonstrated to be very significant for building flexible touch screen and OLED [29-32]. Graphene oxide (GO) is an insulating form of grapheme with an apparent band gap of 3.6 eV. Lee et.al. reported the fabrication of flexible graphene-based thin-film transistors (TFTs) on plastic substrates using graphene oxide (GO) as the gate dielectrics [33]. The dielectric constant of the GO dielectric films was estimated to be 5.0 at room temperature at 1.0 KHz, and no electric-double-layer (EDL) effect was found. Recently, it was reported that GO was a good proton conducting electrolyte in air ambient [34], and extremely large specific gate capacitance can be measured in air ambient due to the EDL effect at low frequency [35]. In this communication, flexible graphene oxide/metal oxide hybrid neuromorphic devices coupled by proton conducting graphene oxide (GO) electrolyte films were fabricated on PET substrates. Boolean logic operation was demonstrated in individual device with multiple input gates. At the same time, dendritic integrations in both spatial and temporal manners were successfully mimicked. What's more, a proof-of-principle visual system for emulating lobula giant motion detector (LGMD) neuron was demonstrated based on such flexible neuromorphic devices.

**Figure 1**a and b shows the schematic diagram of graphene film deposition on Cu foil by chemical vapor deposition (CVD) method, and the roll-to-roll production of conducting graphene film on flexible PET substrate. Basically, the graphene on Cu foil was attached to one piece of thermal release tape by a roll-to-roll machine, which was subsequently immersed in ammonium peroxydisulfate solution for etching Cu foil [29]. Next, the graphene film can be transferred to the PET substrate after a roll-to-roll thermal releasing process at 120 $^{\mathrm{o}}$C. Figure

1c shows the sheet resistance mapping results of flexible graphene/PET sample in the area of 6.0 cm×6.0 cm. An average sheet resistance of ~217 $\Omega \cdot \square^{-1}$ is obtained. The deviation ($\delta$) is determined to be ~7.2 $\Omega \cdot \square^{-1}$. Such results show a small deviation value and good sheet resistance uniformity of the graphene on PET substrate. Then, grapnene oxide (GO) films with a thickness of    500 nm were spin-coated onto the graphene/PET substrates, as shown in the Figure 1d. Then, a 30 nm thick patterned indium-zinc-oxide (IZO) channel layer was deposited on the GO films by radio-frequency (RF) magnetron sputtering with the aid of a nickel shadow mask. The sputtering was performed using an IZO ceramic target. Finally, patterned 100-nm thick Au source/drain and multiple in-plane gate electrodes (1000 μm×150 μm) were deposited through another nickel shadow mask by thermal evaporation method. The detailed descriptions are presented in the experimental section.

**Figure 2**a shows a schematic image of a multi-gate neuron transistor on conducting flexible grapnene/PET substrate. Voltage applied on the in-plane gate electrodes (Au electrodes) can be coupled to the IZO channel layer through two GO gate capacitors in series due to the existence of a bottom conducting graphene layer (bottom gate electrode).  Figure 2b shows a picture of the IZO-based neuromorphic device array on graphene/PET substrate. Figure 2c shows the frequency dependent specific capacitance of the GO film in the frequency range of 1.0 Hz to 1.0 MHz. The capacitance was measured with a graphene/GO/Au sandwiched structure (inset in Figure2c). A maximum specific capacitance of ~4.0 μF/cm$^2$ is obtained at 1.0 Hz due to the formation of the EDL at the interfaces, and the specific capacitance value decreases with increasing frequency. The absolute value of leakage current of GO film is lower than ~4.0 nA, indicating the good insulativity of the GO film for neuron transistor application. Figure 2d shows the output characteristics ($I_{DS}$ vs $V_{DS}$) of the IZO-based neuron transistor with $V_{GS}$ varied from -0.5 to 1.5 V in 0.25 V steps. At low $V_{DS}$, $I_{DS}$ increases linearly with $V_{DS}$, indicating that the device has a good Ohmic contact between IZO and Au source/drain electrodes. At higher $V_{DS}$, the drain current gradually approaches to a saturated

value. Such results indicate that the proposed device is a typical n-type field-effect transistor.

Figure 2e shows the transfer curves ($V_{DS}$=1.5 V) of the flexible neuron transistors with different bending times. The gate voltage is swept from -1.5 V to 1.5 V and then back. An anticlockwise hysteresis loop of ~0.2 V is observed, which is likely due to the existence of mobile protons in the GO film[36, 37]. No appreciable degradation in performance can be observed even after bending for 1000 times. A high current ON/OFF ratio of ~2×10^6 and a low subthreshold slope ~100 mV/dec are obtained. The field-effect mobility ($\mu$) of the transistors for repeated bending test can be extracted from the device equation at the saturation region ($V_{DS}$>$V_{GS}$−$V_{TH}$): $I_{DS}$=($WC_i\mu/2L$)($V_{GS}$−$V_{TH}$)$^2$, where L is the channel length, W is the channel width, and $C_i$ is the specific capacitance of the GO film. The inset figure shows the photograph of the bended device array. As shown in Figure 2f, the mean value of $\mu$ is estimated to ~15.5 cm$^2$/V·s with a standard deviation of ~1.1 cm$^2$/V·s. Figure 2f also shows the influence of bending on the threshold voltage of the flexible neuron transistor. The threshold voltage shift is found to be less than 0.1 V even after 1000 times bending. These results suggest the good mechanical flexibility and stability of the graphene/metal oxide hybrid neuron transistors.

**Figure 3**a shows the transfer curves of the neuron transistors with two in-plane gate inputs (G1 and G2) when $V_{GS1}$ sweeps from -1.5 to 1.5 V with $V_{GS2}$ fixed at -1.0 and 1.0 V, respectively. When $V_{GS2}$=-1.0 V, only very low drain current <2.0 nA can be measured. When $V_{GS2}$=1.0 V, the high resistance to low resistance to transition of the IZO channel layer can be observed when VGS1 sweeps from negative voltage to positive voltage. As shown in Figure 3b, the majority protons in the GO film will migrate to the GO/G2 gate electrode interface when $V_{GS2}$=-1.0 V. When $V_{GS1}$ sweeps from -1.5 to 1.5 V, few protons can migrate to the GO/IZO channel interface. In this case, few electrons can be electrostatically coupled in the IZO channel layer, which will result in a low drain current. On the contrary, when $V_{GS2}$=1.0 V and $V_{GS1}$ sweeps from -1.5 to 1.5 V, abundant protons will migrate to the GO/IZO channel

layer interface, and extremely strong EDL capacitive coupling can result a high conductivity in the IZO channel layer. In a conclusion, the ON state (>100 μA) of the IZO channel can only be achieved when both of the two gate inputs were ≥1.0 V. Such phenomenon is very similar to the 'AND' logic. If we set the input voltage of -1.0 V to be logic '0' and 1.0 V to be logic '1', an 'AND' logic can be achieved as shown in Figure 3c [38]. Two square wave sequences with different periods (the former is twice than the latter) were applied on G1 and G2, respectively, thus the two bits binary inputs of '00', '01', '10' and '11' can be obtained sequentially. As shown in Figure 3c, the ON state of IZO channel can only be achieved when the input signals are '11', which indicate the 'AND' logic. As the ON/OFF ratio is higher than $10^6$, thus the operation of this logic gate is very robust. A simple resistor-loaded inverter was also demonstrated by connecting the IZO-based neuron transistor in series with a 2.0 MΩ resistor, as shown in the inset of Figure 3d. The voltage transfer characteristics at $V_{DD}$=0.5, 1.0, 1.5 and 2.0 V are shown in Figure 3d. The voltage gain ($\partial V_{OUT}/\partial V_{IN}$) is dependent on $V_{DD}$, and the maximum gain of 3.0, 6.2, 8.7 and 10 can be obtained at $V_{DD}$=0.5, 1.0, 1.5 and 2.0 V, respectively. These results clearly suggested that such flexible neuromorphic devices can greatly simplify the design for logic circuit applications.

Next, synaptic/neuronal emulations in individual flexible neuromorphic device will be discussed. To mimic the behaviors of the biological synapse, voltage pulses applied on the gate electrodes are regarded as the presynaptic spikes, and the induced channel current is defined as the postsynaptic current [14-18]. **Figure 4**a shows a typical excitatory postsynaptic current (EPSC) measured from the flexible neuron transistors with a reading voltage of $V_{DS}$=0.1 V. The amplitude and duration of presynaptic spike are 0.5 V and 10 ms, respectively. The EPSC dramatically increases to a maximum value of ~97 nA at the end of the spike and decay to the baseline within a time window of tens millisecond. The slow decay of the EPSC indicates the electrostatic interactions between protons in the GO film and electrons in the IZO channel. More protons would be triggered and accumulated with a longer spike duration, which would augment the EPSC as a result. The decay of EPSC can be fitted well by the following equation:

$$A(t) = A_0 + A \cdot \exp[-(\frac{t}{\tau})^{\beta}] \qquad (1)$$

where $A_0$, $A$, $\tau$ and $\beta$ are the channel current measured at $V_{GS}$=0 V, the EPSCamplitude, time constant and material-dependent index, respectively. Figure 4b shows a two layer model of hierarchical parallel processing in dendritic trees [39, 40]. Presynaptic spiking inputs are integrated locally in dendritic subunits, each of which computes the sum of its local inputs nonlinearity. Dendritic integration plays an important role in information transformation including addition of nonsimultaneous unitary events (temporal summation) and addition of unitary events occurring simultaneously in separate regions of the dendrite (spatial summation)[41]. First, a form of temporal summation is illustrated in Figure 4c [42]. Two temporal correlated spikes (V1, V2, 0.2 V, 10 ms)) were applied on G1 and G2, respectively. If the zero time is defined at V2 just ended, the EPSC amplitude measured at zero time is highly dependent on the time interval between the two spikes. For example, when V1 is triggered before V2 with a time interval of 40 ms (ΔT=-40 ms), the EPSC amplitude measured at zero time is 41 nA. When the V1 is triggered synchronously with the V2, the EPSC amplitude is 120 nA. When the V1 is triggered after the V2 with a time interval of 40 ms (ΔT=40 ms), the EPSC amplitude is 16 nA. Figure 4d illustrates a dynamic logic established by the spatiotemporal correlated spikes, which is unsymmetrical with increasing |ΔT|. When ΔT>0, the EPSC amplitude measured at zero time is equal to the amplitude triggered individually by V2. When ΔT≤0, the residual protons triggered by V1 can augment the followed EPSC. In that case, EPSC measured at zero time is larger than the EPSC triggered individually by V2. When ΔT decreases further, the increase in the EPSC amplitude at zero time by V1 will be gradually less significant.

The spatial summation of the flexible neuron transistor will be also investigated. As shown in **Figure 5**a, the EPSCs were triggered by two presynaptic spikes (0.5 V, 10 ms) applied on two in-plane gate electrodes (G1 and G2) with the same electrode areas, respectively. The triggered EPSC is measured at a reading voltage $V_{DS}$=0.1 V. The amplitude of the first EPSC is ~93 nA, which is almost equal to the second one. Then the two presynaptic spikes are triggered simultaneously. The amplitudes of the two presynaptic spikes are systematically changed from 0.2 to 1.4 V. Figure 5b shows the spatial summation result which is plotted as 2D surface. If the threshold current is set to be 750 nA, the summation EPSC by the two spikes can exceed such value only when both of the two presynaptic voltages are high enough (e. g. ≥1.0 V). For example, the EPSC amplitude is ~510 nA and ~830 nA for (1.2 V, 0.2 V) and (1.2 V, 1.0 V), respectively. As shown in Figure 5c, the two presynaptic spikes (0.5, 10 ms) were applied on two gate patterns (G1 and G2') with different electrode areas. The area of Gate 2' is ~2.5 folds larger than that of the Gate 1. The second EPSC amplitude of ~350 nA is much larger than the first one. The voltage applied on Gate 1 and Gate 2' are systematically changed from 0.2 to 1.4 V, and the spatial summation result is plotted as 2D surface as shown in Figure 5d. In this case, the summation of EPSC by the two spikes can exceed the threshold value so long as V2 is high enough (e. g. ≥1.0 V).

The two types of spatial summation can be analogous to the logic 'AND' and 'YES$_{V2}$', respectively. The input voltage of 0 and 1.0 V are defined as '0' and '1', respectively. The duration of the spiking inputs are 10 ms. The integrated EPSC amplitude is defined as the output and the threshold is set to 750 nA. As shown in left panel of Figure 5e, the binary inputs of '00', '01', '10' and '11' were applied on Gate 1 and Gate 2'. Only when input signals are '11', the EPSC amplitude is larger than the EPSC threshold, which indicates the 'AND' logic. As shown in the right panel of Figure 5e, as long as input $V_2$ is '1', the EPSC amplitude is larger than the EPSC threshold, which indicates the 'YES$_{V2}$' logic. Figure 5f shows the truth tables and the ideal summation surfaces for the two logics. Such logics can have important implications in capturing the computing powers in neuronal system where the nonlinear and analog mechanisms are predominant [23, 43]. What's more, in a neural network, the synaptic weight distribution is varied dependent on a number of factors such as the synapse location [44, 45]. Therefore, the electrode area can be regarded as a key factor for designing the synaptic weight distribution in the neuromorphic system based on the proposed artificial neuron.

In nature, the ability of an animal to detect approaching objects is quite essential to prevent collisions and avoid capture by predators [46, 47]. The lobula giant movement detector (LGMD) neuron is part of a neural circuit thought to be involved in the generation of escape behaviors. Approaching objects are distinguished by the LGMD using the increasing speed of edge movement and increasing length of the edges [46]. Previously, emulation of LGMD behaviors was realized based on large-scale integration circuit with high energy consumption [48, 49]. Here, we propose a proof-of-principle neuromorphic module based on our neuromorphic devices for LGMD neuron behaviors emulation. As shown in **Figure 6**a, the 20×20 photoreceptor array ($P_{11}$, $P_{12}$, …) is connected to the multiple in-plane gate arrays ($G_{11}$, $G_{12}$, …) of the flexible neuromorphic transistor through a processing circuit. A square object moves in front of the array. Both the object and photoreceptor array are paralleled with the *xoy* plane. The center coordinates of the object and the array are $(0, 0, z_o)$ and $(0, 0, z_a)$, respectively. The gray dashed box denotes the image of the object in the photoreceptor array. The edge movement thus can be detected by comparing the difference between the successive images. The excitatory stimulus (0.5 V, 1.0 ms) can be triggered only when the object edge is detected by a photoreceptor. The excitatory stimuli are then send to corresponding presynaptic terminals (gates) of the neuron transistor (See details in S1 to S4, supporting information).

In our simulation, the inputs to the network are a series of computer-generated images of the moving object with a fixed *l/v*, where *l* is the object's half-size and *v* is approach speed. The simulated time is set to be one per millisecond. Each image is mapped onto the photoreceptors array. The output of the flexible neuron transistor is calculated based on the Eq. 1. The parameters of $A_0$, $A$, $\tau$ and $\beta$ were set to be 10 nA, 37.5 nA, 1.06 ms and 0.35, respectively. These parameters are estimated from the experimental data of EPSCs. The weights of all the presynaptic terminals were set equally. Firstly, when the object moved toward the photoreceptor array with $l/v$=50 ms, the response of the neuron transistor is shown in Figure 6b. The response of the neuron transistor increases dramatically when the object is approaching as shown in Figure 6b. The maximum value of ~2.2 µA is obtained at 28 ms before collision. At the beginning, the angular size of the object is smallest, and few of the photoreceptors can detect the changes of the edges. When the object is looming, the edges length becomes larger and the edge movement becomes faster. Therefore, more excitatory stimuli can be triggered with an increased frequency. However, when the object is moving away at the same speed, the neuron transistor responses intensively at the beginning, and decays gradually after that as shown in Figure 6c. The maximum value of ~1.7 µA is obtained at 8.0 ms after the first movement. Finally, Figure 6d shows the object moved in the *xoy* plane from side to side of the array. For the translation motion, the speed of edge movement and the length of the edge are constant. The speed of edge movement is set to be 0.1 photoreceptor/ms, and the corresponding angular size of the object is fixed at 93$^\text{o}$. In this case, the excitatory stimuli with a constant frequency (100 Hz) can be triggered, and a slow increase of EPSC with maximum value of ~1.2 µA is achieved. Such results indicate that such visual system model can respond selectively to objects approaching on a collision course. Such visual system based on the proposed neuron transistors could be further utilized for collision avoidance and other intelligent behaviors.

In summary, flexible graphene oxide/metal oxide hybrid neuromorphic transistors gated by proton conducting graphene oxide electrolyte films were fabricated on PET substrates. Such flexible neuromorphic devices were experimental demonstrated with good mechanical flexibility and electrical performance. More importantly, dendritic integrations in both spatial and temporal modes were successfully mimicked, and spatiotemporal correlated logics were realized. A proof-of-principle visual system model for emulating LGMD neuron was demonstrated, which responds selectively to objects approaching on a collision course. Our results are of great interest for conformable bionic sensors, brain-computer interface and neuromorphic cognitive systems.

**Experimental Section**

*Preparation and Characterizations of monolayer graphene PET substrate:* The graphene PET film was prepared through a CVD and following transfer process. In a typical fabrication process, copper foil (Cu, 99.99%) in size of $10 \times 20$ cm$^2$ was loaded in a tube furnace which was vacuumed to 10 Pa and then heated to 1000 $^o$C within 2 hours with 8 sccm H$_2$ gas. After annealing Cu foil at 1000 $^o$C for additional 30 minutes, 24 sccm CH$_4$ was introduced into the tube furnace for CVD growth of graphene film for 30 minutes. The graphene on Cu foil can be collected after cooling the furnace to room temperature. The graphene film was then transferred to a transparent PET substrate with the help of thermal release tape. Basically, the graphene on Cu foil was attached to one piece of thermal release tape by a roll-to-roll machine, which was subsequently immersed in 0.1 M ammonium peroxydisulfate solution for etching Cu foil. Then, the graphene film can be transferred to the PET substrate after a roll-to-roll thermal releasing process at 120 $^o$C [18]. The final graphene PET was obtained after HNO$_3$ treatment and DI water washing. The mean sheet resistance of the graphene PET with single layer of graphene is measured as ~217 $\Omega \cdot \square^{-1}$ by the four-probe (NAPSON CRESBOX) measurement.

*Preparation and Characterizations of Graphene Oxide (GO) films:* Firstly, 1.5 g natural graphite flakes and 1.8 g KNO$_3$ was added into 69 mL of concentrated H$_2$SO$_4$ (98%) under stirring. After a few minutes, 12 g KMnO$_4$ was added slowly. The mixture was then heated to 40 °C and stirred for 6 hours. Subsequently, 120 mL water was added under vigorous stirring, resulting in a quick rise of the temperature to ~80 °C. The slurry was further stirred at this temperature for another 30 min. Afterwards, 300 mL water and 12 mL $H_2O_2$ solution (30 wt.%) were added in sequence to dissolve insoluble manganese species. The resulting graphite oxide suspension was washed repeatedly by plenty of water until the solution pH reached a constant value of ~5.0. Complete delamitation of graphite oxide into GO was achieved by ultrasonic treatment. Finally, the brown, homogeneous colloidal suspension of GO was obtained. X-ray photoelectron spectroscopy (XPS) measurements of GO films were carried out on AXIS UTLTRA DLD. For XPS measurement, GO solution were also spin-coated on a polished Si (100) wafer. The thickness of the GO films is measured by Stylus Profiler (Dektak150, Veeco).

*Fabrication of Neuron Transistors:* Flexible GO coupled IZO neuron transistors on graphene conducting substrates were fabricated at room temperature. Firstly, GO suspensions (~6 mg/mL) were spin-coated onto graphene PET substrates and dried at 50 °C for 2 h. The thickness of the GO film is estimated to 1.2 μm by Stylus Profiler. Then, a 30 nm thick patterned indium-zinc-oxide (IZO) channel layer was deposited on the GO films by radio-frequency (RF) magnetron sputtering with the aid of a nickel shadow mask. The sputtering was performed using an IZO ceramic target with a RF power of 100 W and a working pressure of 0.5 Pa. The channel width and length were 1000 μm and 80 μm, respectively. Finally, patterned 150-nm thick IZO source/drain and gate electrodes (1000 μm×150 μm) were deposited through another nickel shadow mask by thermal evaporation method.

*Device Electrical Characterizations*: The frequency dependent capacitances of GO films were characterized by a Solartron 1260A Impedance/Gain-Phase Analyzer. The electrical measurements of the neuron transistors were performed on a semiconductor parameter characterization system (Keithley 4200 SCS) at room temperature with a relative humidity (RH) of 50 %.

*Acknowledgements*

This work was supported in part by the National Science Foundation for Distinguished Young Scholars of China (Grant No. 61425020), and in part by the Zhejiang Provincial Natural Science Fund (LR13F040001).

*Supporting Information*

Supporting Information is available online from Wiley InterScience or from the author.

Received: ((will be filled in by the editorial staff))
Revised: ((will be filled in by the editorial staff))
Published online: ((will be filled in by the editorial staff))

**Figure captions**

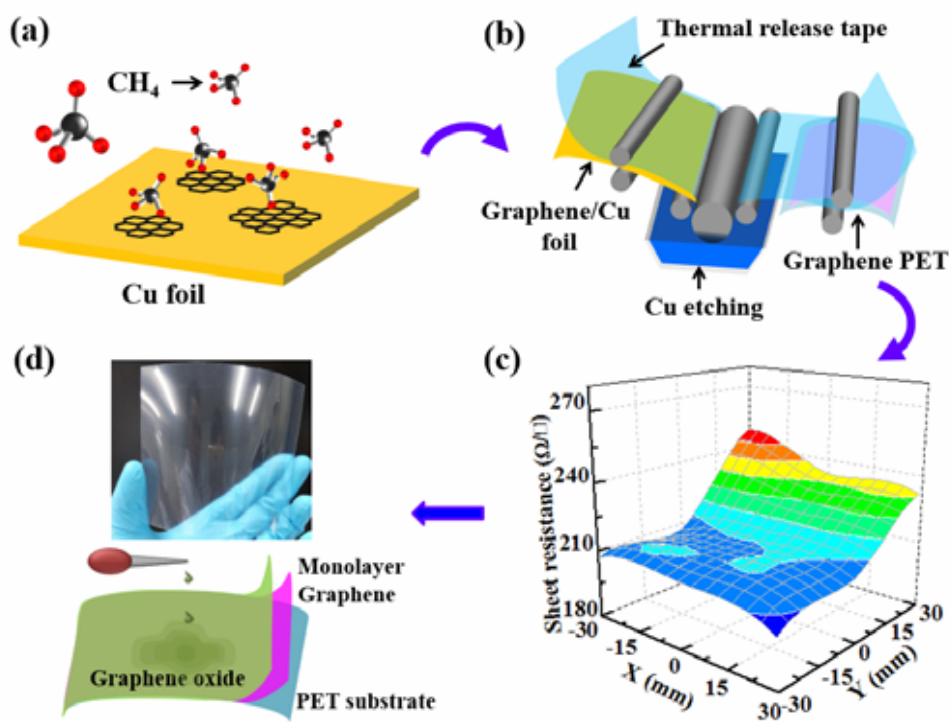

**Figure 1**. a) Schematic diagram of graphene film deposition on Cu foil by chemical vapor deposition (CVD) method. b) Schematic diagram of the roll-to-roll production of conducting graphene film on flexible PET substrate. c) Resistance mapping of sheet resistance for the flexible graphene/PET. The mapped area is ~6.0 cm×6.0 cm. d) Picture of the flexible graphene/PET sample (10 cm×15 cm), and schematic image of the grapnene oxide (GO) films (~500 nm) deposition on the graphene/PET substrate by spin coating method.

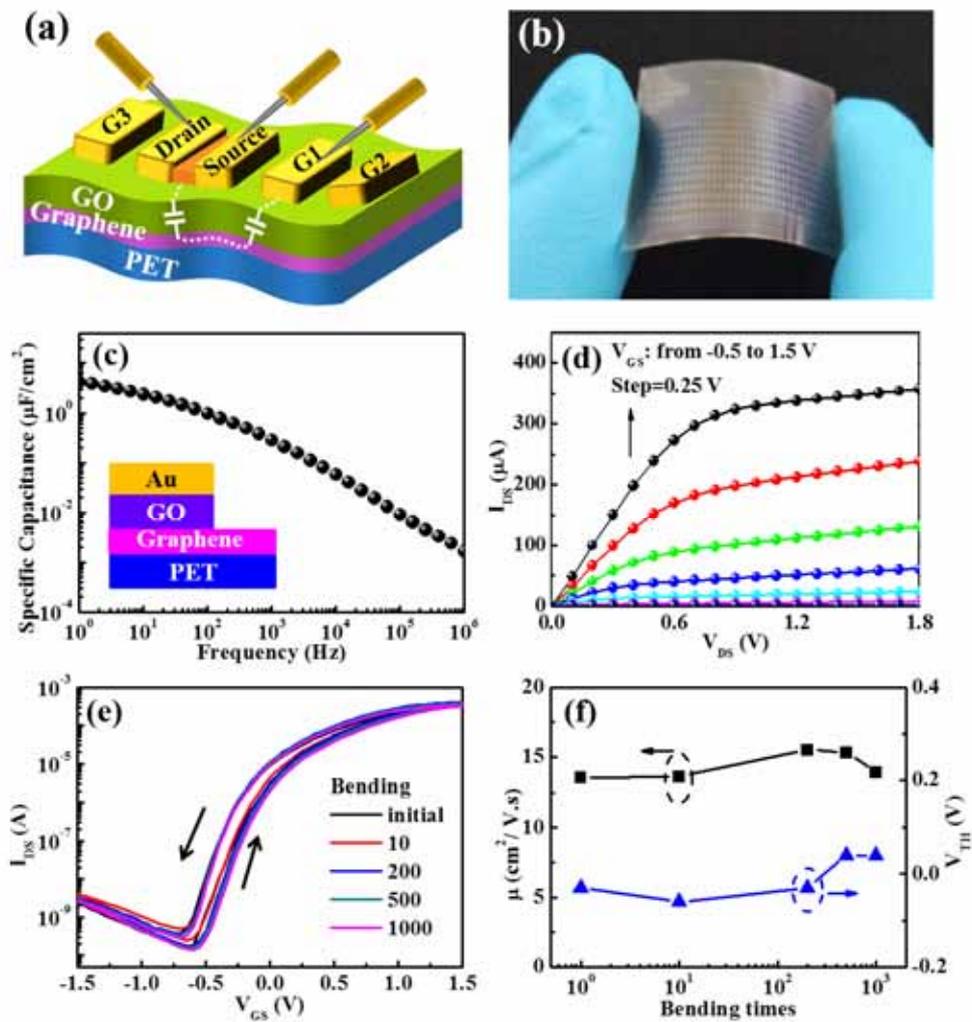

**Figure 2**. a) Schematic image of the multi-gate GO/IZO hybrid neuron transistors on flexible conducting grapnene/PET substrate. b) Picture of the neuron transistor array gated by GO on graphene/PET substrate. c) Frequency-dependent specific capacitance of the GO film. Inset: Graphene/GO/Au sandwich structure for capacitance and gate leakage current measurements. d) The output characteristics ($I_{DS}$ vs $V_{DS}$) of the flexible neuron transistor on grapnene/PET substrate. e) Influence of bending on the transfer curves measured at $V_{DS}$ =1.5 V. f) Field-effect mobility and threshold voltage of the neuron transistor measured after different bending cycles.

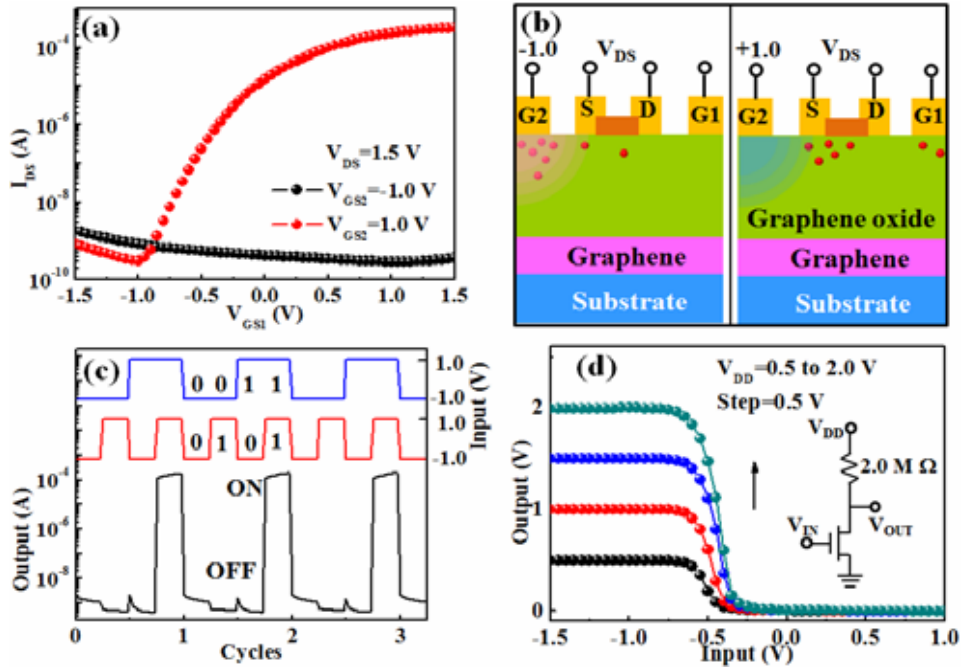

**Figure 3**. a) Transfer curves ($I_{DS}$-$V_{GS1}$) for the flexible GO/IZO hybrid neuron transistors with the $V_{GS2}$ biased at -1.0 and 1.0 V, respectively. b) The schematic diagrams illustrate the proton profile under $V_{GS2}$ of -1.0 V (left panel) and 1.0 V (right panel), respectively. c) Input-output characteristics of the 'AND' logic from the transistors. d) Static behaviors of a resistor loaded inverter with different $V_{DD}$ ranged from 0.5 to 2.0 V in 0.5 V steps.

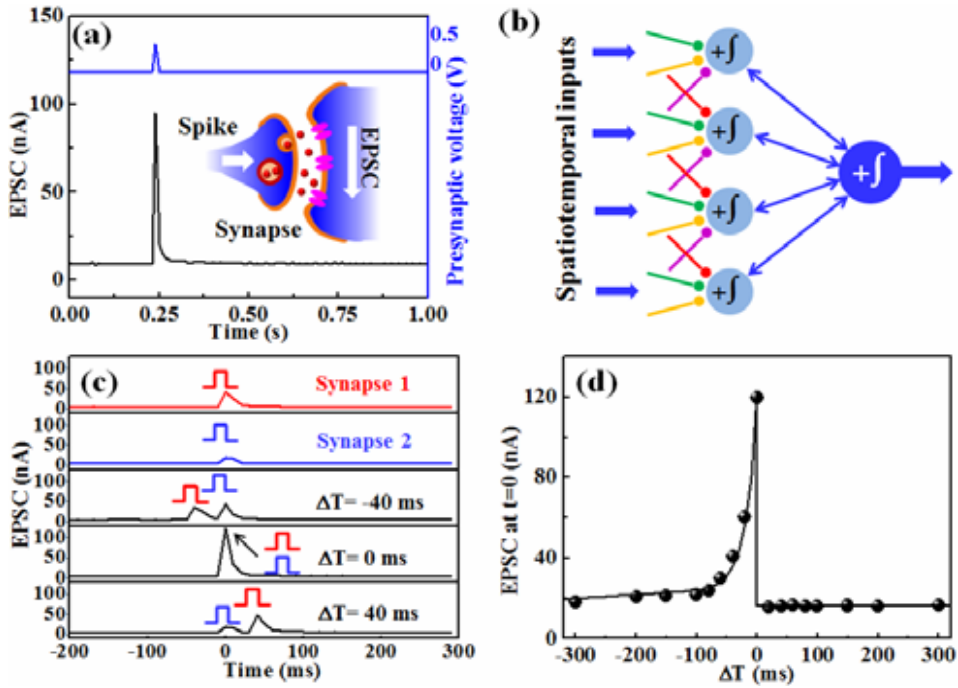

**Figure 4**. a) Typical EPSC recorded in response to a presynaptic spike (0.5 V, 10 ms) applied on the gate with $V_{DS}$=0.1 V. b) The schematic diagram illustrated a two layer model of hierarchical parallel processing in dendritic trees. c) The integrated EPSCs recorded in response to two temporal correlated inputs (0.2 V, 10 ms) with different time intervals of -40, 0 and 40 ms, respectively. The EPSCs triggered by each input is also shown in the figure. d) The dynamic logic established by the two spatiotemporal correlated spikes. The EPSC amplitudes were measured at t=0.

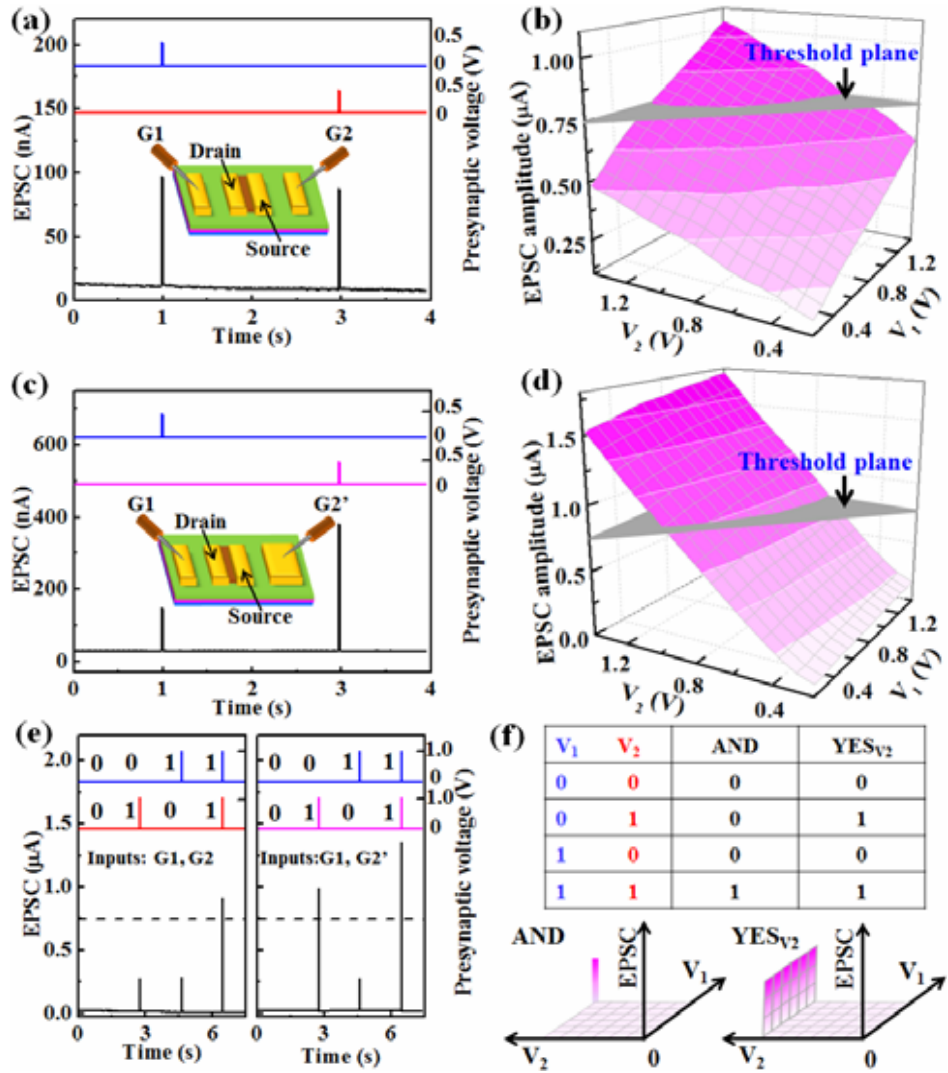

**Figure 5**. EPSCs triggered by two presynaptic spikes (0.5 V, 10 ms) applied on a) G1 and G2, and c) G1 and G2', respectively. b) and d) The corresponding spatial summation results plotted as 2D surface, respectively. The two presynaptic spikes ($V_1$ and $V_2$) changed from 0.2 to 1.4 V with duration of 10 ms. e) Input-output characteristics of the 'AND' and 'YES$_{V2}$' logic from two types of presynaptic input terminals. f) The truth table and ideal summation surface for the two logics.

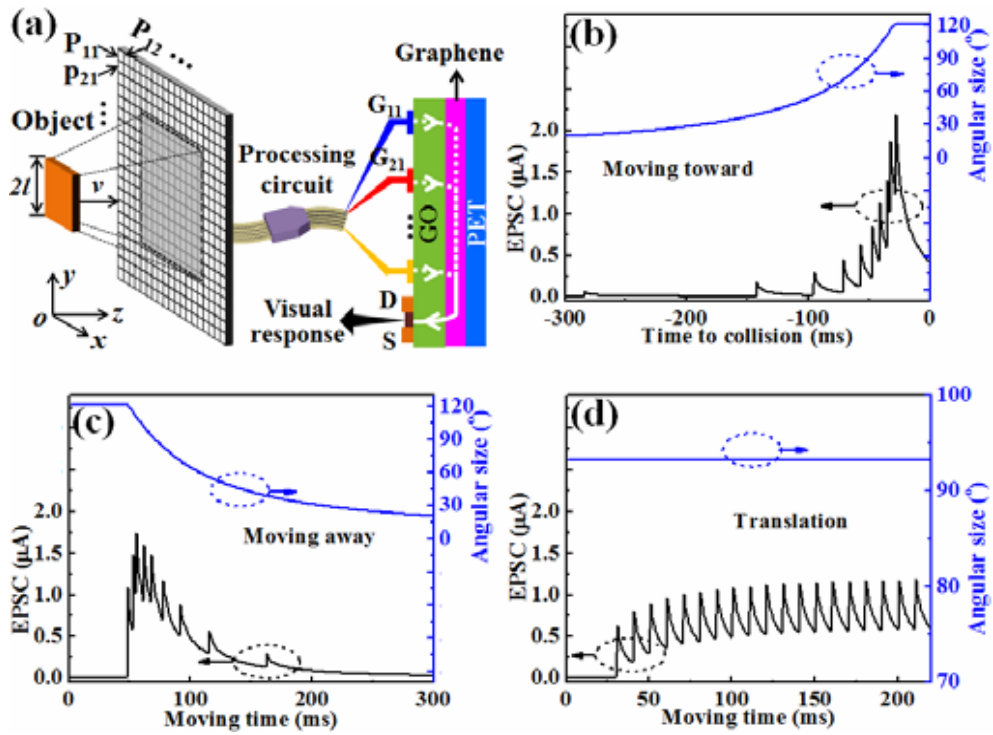

**Figure 6**. a) The proof-of-principle visual system for emulating the LGMD neuron based on our GO/IZO hybrid neuron transistors with multiple gate arrays. The EPSC output of the visual system model recorded in response to an object when it is b) moving toward, c) moving away and d) moving paralleled to the photoreceptor array, respectively.

**The table of contents entry.** Flexible metal oxide/graphene oxide hybrid multi-gate neuron transistors were fabricated on flexible graphene substrates. Dendritic integrations in both spatial and temporal modes were successfully emulated, and spatiotemporal correlated logics were obtained. A proof-of-principle visual system model for emulating lobula giant motion detector neuronwas investigated. Our results are of great interest for flexible neuromorphic cognitive systems.

C. J. Wan, W. Wang, L. Q. Zhu, Y. H. Liu, P. Feng, Z. P. Liu, Y. Shi, Q. Wan

## Flexible Metal Oxide/Graphene Oxide Hybrid Neuromorphic Devices on Flexible Conducting Graphene Substrates

ToC figure ((55 mm broad, 50 mm high, or 110 mm broad, 20 mm high))

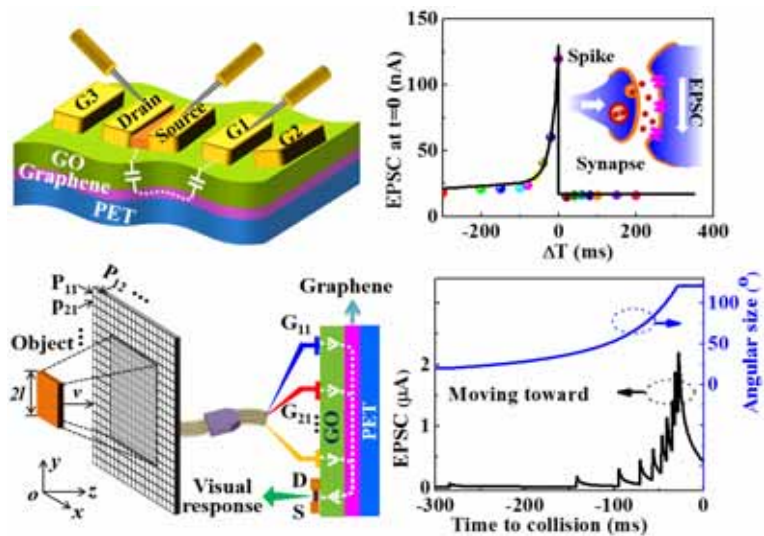

# Flexible Metal Oxide/Graphene Oxide Hybrid Neuromorphic Devices on Flexible Conducting Graphene Substrates

Chang Jin Wan [1,2], Wei Wang [2], Li Qiang Zhu [2], Yang Hui Liu [2], Ping Feng [1], Zhao Ping Liu [2], Yi Shi [1*], and Qing Wan, [1,2,*].

## S1 When the object is approaching to the array.

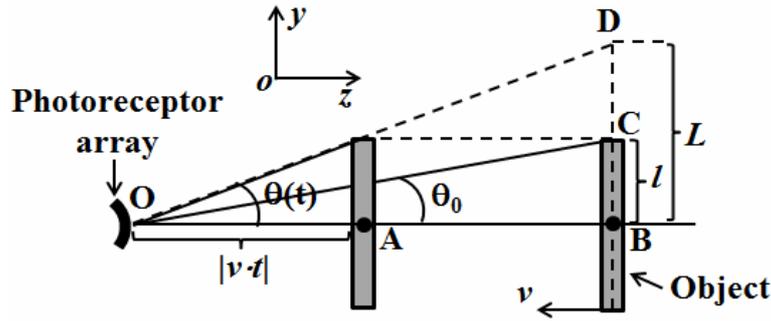

Figure S1. The schematic diagram illustrates the geometrical relationship between the object and the photoreceptor in yoz plane. The object is moving toward the photoreceptor array from B to A at a speed of *v*. At the beginning, the angular size of the object at the array is $2 \cdot \theta_0$, and the edge length in the array is equivalent to $2 \cdot n_0$ photoreceptors ($n_0$ is an integer). The size of the object is $2 \cdot l$. The size of photoreceptor array is N×N (N=2n, n=1, 2, …).

As shown in Figure S1, the angular size of object is $2 \cdot \theta(t)$. The angular size is given by trigonometry that $\tan\theta(t)=l/|v \cdot t|$, where *v* and *t* are motion speed and time to collision (t<0), respectively. In that case, the edge length of the object (2·L) in the photoreceptor array is determined by $2 \cdot n_0 \cdot \tan\theta(t)/\tan\theta_0$. Therefore, L can be estimated by the following equation:

$$L = \frac{n_0 \cdot (l / v)}{|t| \cdot \tan \theta_0} \qquad (S1)$$

When the object was moving toward the array, the edges of the object in the array moved from the center to the periphery. Figure S2 shows three cases during the looming of the object, where the size of the array is 10×10 (N=10). As shown in Figure S2a, the edges of the object were detected by $P_{55}$, $P_{65}$, $P_{56}$ and $P_{66}$. As shown in figure S2b, the edges were detected by $P_{44}$, $P_{54}$, $P_{64}$, $P_{74}$, $P_{45}$, $P_{75}$, $P_{46}$, $P_{76}$, $P_{47}$, $P_{57}$, $P_{67}$ and $P_{77}$. As shown in figure S2c, the edges were detected by $P_{33}$, $P_{43}$, $P_{53}$, $P_{63}$, $P_{73}$, $P_{83}$, $P_{34}$, $P_{84}$, $P_{35}$, $P_{85}$, $P_{36}$, $P_{86}$, $P_{37}$, $P_{87}$, $P_{38}$, $P_{48}$, $P_{58}$, $P_{68}$, $P_{78}$ and $P_{88}$. When the object edge was detected by a photoreceptor, the excitatory stimulus (0.5 V, 1 ms) would be triggered and send to the corresponding gates ($G_{11}$, $G_{12}$…, as shown in Fig. 6a in the manuscript) of the graphene/metal oxide hybrid neuron transistors.

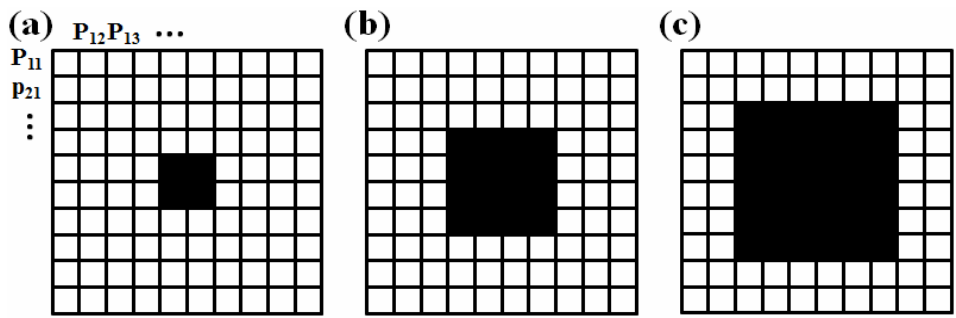

Figure S2. (a) to (c) When the object is moving toward the array, the edges will move in the array from the center area to the periphery.

In that case, when the edge length in the array (2·L) is integer multiples of the size photoreceptor size, the excitatory spikes can be generated. Therefore the triggering time (T) of spikes can be estimated by the following equation:

$$T = -\frac{l/v}{n \cdot \tan\theta_0} \qquad (S2)$$

where $n$=1, 2,…N/2.

The number of the excitatory photoreceptors for each k is 4(2n-1) deduced from the Figure S2.

## S2 When the object is moving away from the array.

On the contrary, when the object is moving away from the array, the object edges are moving from the periphery to the center. Therefore the triggering time (T) of spikes can be estimated by the following equation:

$$T = \frac{l/v}{n \cdot \tan\theta_0} \qquad (S2)$$

where $n$=N/2, N/2-1,…1.

And the number of the excitatory photoreceptors for each k is 4(2n-1), too.

**S3 When the object is moving paralleled with the array.**

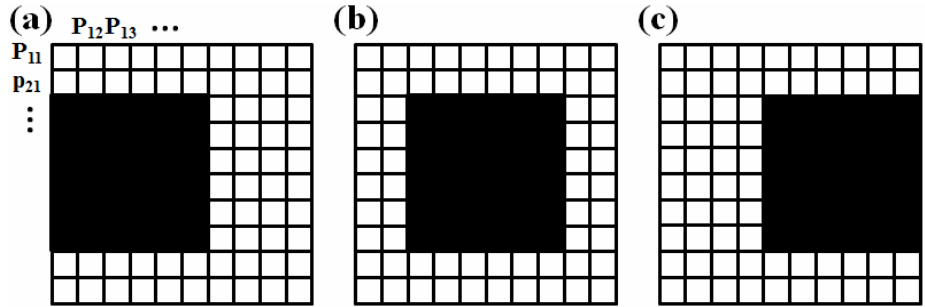

Figure S3 (a) to (c) show three cases during the object is moving rightward the photoreceptor array.

When the object is moving rightward (or leftward), the edge length in the array is constant ($2 \cdot n_0$), as shown in Figure S3. And the frequency of the excitatory spikes is fixed and dependent on the motion speed. If the size of a photoreceptor is $a$, the time interval between the excitatory spikes is $a/v$.

**S4 The parameters for the simulations.**

The parameters for the three cases, i. e. approaching, moved away and translation, were listed in table SI, SII and SIII, respectively.

Table SI

| $l/v$ | $n$ | $\theta_0$ | $n_0$ |
|---|---|---|---|
| 50 ms | 10 | 10$^\text{o}$ | 1 |

Table SII

| $l/v$ | $n$ | $\theta_0$ | $n_0$ |
|---|---|---|---|
| 50 ms | 10 | 10$^\text{o}$ | 1 |

Table SIII

| $a/v$ | $n_0$ |
|---|---|
| 10 ms | 6 |